\documentclass[amssymb,aps]{revtex4}
\begin{document}

\title{Supersymmetric Non-local Gas Equation}  

\author{Ashok Das$^{a}$ and Z. Popowicz$^{b}$}
\affiliation{$^{a}$ Department of Physics and Astronomy,
University of Rochester,
Rochester, New York 14627-0171, USA}
\affiliation{$^{b}$ Institute of Theoretical Physics,
University of Wroc\l aw, 
pl. M. Borna 9, 50 -205 Wroc\l aw, Poland} 

\bigskip

\begin{abstract}
In this paper we study systematically the question of
supersymmetrization of the non-local gas equation. We obtain both the
$N=1$ and the $N=2$ supersymmetric generalizations of the system which
are integrable. We show that both the systems are
bi-Hamiltonian. While the $N=1$ supersymmetrization allows the
hierarchy of equations to be extended to negative orders (local
equations),  we argue that this is not the case for the $N=2$
supersymmetrization. In the bosonic limit, however, the $N=2$ system
of equations lead to a new coupled integrable system of equations.

\end{abstract}


\maketitle

\section{Introduction}

The classical isentropic gas equations \cite{whitham}
\begin{eqnarray}
u_{t} + uu_{x} + \frac{1}{v}\ P_{x} & = & 0,\nonumber\\
v_{t} + \left(vu\right)_{x} & = & 0,
\end{eqnarray}
where $u,v$ denote respectively the velocity and the density of the
gas, are known to constitute an interesting class of dispersionless
integrable systems when the pressure is a monomial function of the
density. For example, $P = v^{\gamma},\gamma \neq 0,1$ corresponds to
the polytropic gas, while $P = - \frac{1}{v}$ describes the Chaplygin
gas \cite{brunelli,nutku}. Both these systems are of hydrodynamic
type \cite{dubrovin}, integrable and a lot
is known about the properties of these systems. Recently, it was shown
that the gas equation for the case $P = - \frac{1}{2}
\left(\partial^{-1}v\right)^{2}$ (called the non-local gas
equation) is also integrable and has a very rich algebraic
structure \cite{brunelli1,pavlov}. This system of equations arises in
astrophysical models of dark matter \cite{gurevich}. In this paper, we
will study the supersymmetrization of
this system maintaining the integrability aspects of the model.

It is worth recalling that supersymmetrization of dispersionless
systems is at best poorly understood at the present \cite{das,das1}.
For example, in
the case of polytropic gas, an integrable supersymmetric hierarchy has
been obtained only for $N=1$ supersymmetrization besides the
``trivial'' susy-B supersymmetrization \cite{mathieu,becker} and the
supersymmetrization of the Chaplygin gas \cite{jackiw} resembles a susy-B
symmetrization \cite{das1}. Even in the case
of the $N=1$ supersymmetric polytropic gas (which is integrable), it
is not known if it is a bi-Hamiltonian system. In contrast, we will
show that integrable $N=1$ and $N=2$ supersymmetrizations are possible
for the system of non-local gas dynamics. Furthermore, the $N=1$
supersymmetric system possesses two Hamiltonian structures which are
compatible so that it is truly a bi-Hamiltonian system and possesses
all the rich algebraic structures of its bosonic counterpart. The
$N=2$ supersymmetrization, on the other hand, leads to a new system
of coupled integrable equations in the bosonic limit.

The paper is organized as follows. In section {\bf II}, we briefly
recapitulate the essential features of the bosonic system and present
results on the $N=1$ supersymmetric generalization that is
integrable. We obtain the bi-Hamiltonian structure, the Casimir
functionals as well as the conserved charges (local and non-local) of
the system. In section {\bf III}, we present the essential results on
the $N=2$ supersymmetrization of the system preserving
integrability. We obtain the bi-Hamiltonian structure and argue that
the second structure has no Casimir functional so that the system of
equations cannot be extended to the negative orders. We present a
brief summary of our results in section {\bf IV}.

\section{$N=1$ Supersmmetrization}

The non-local gas equation is described by the system of equations
\begin{eqnarray}
u_{t} & = & - uu_{x} + \left(\partial^{-1}v\right),\nonumber\\
v_{t} & = & - \left(uv\right)_{x}.\label{gasequation}
\end{eqnarray}
This system of equations is known to be integrable and a lot of the
algebraic properties for the system are well known.  For example, it
is known that the system of equations (\ref{gasequation}) is a
bi-Hamiltonian system with the two compatible Hamiltonian structures
described by
\begin{equation}
{\cal D}_{1} = \left(\begin{array}{rr}
0 & - \partial\\
-\partial & 0
\end{array}\right),\qquad {\cal D}_{2} = \left(\begin{array}{cc}
\partial^{-1} & - u_{x}\\
u_{x} & - (v\partial + \partial v)
\end{array}\right).
\end{equation}
The conserved charges satisfying the recursion relation can be
constructed recursively and these charges are in involution by
construction making the system integrable.
Furthermore, the Hamiltonian structure ${\cal D}_{1}$ has three
Casimir functionals (conserved charges whose gradients are annihilated
by the Hamiltonian structures)
\begin{eqnarray}
H_{1} & = & \int \mathrm{d}x\ v,\nonumber\\
H_{1}^{(1)} & = & - \int \mathrm{d}x\ u_{x} \rightarrow 0,\nonumber\\
H_{1}^{(2)} & = & \int \mathrm{d}x\ u,
\end{eqnarray}
while the second Hamiltonian structure has a single Casimir functional
\begin{equation}
H_{-1} = 2 \int \mathrm{d}x\ \left(v - \frac{1}{2}
u_{x}^{2}\right)^{\frac{1}{2}}.
\end{equation}
The existence of the Casimir functionals allows the hierarchy of flows
to be extended to the negative order using the recursion relation and
these are local equations unlike (\ref{gasequation}). In addition to
the charges that are recursively constructed, the model also possesses
two series of conserved charges whose gradients are not related by the
recursion operator (and, therefore, are not in involution with the
infinitely many charges which are in involution)
\begin{equation}
G_{n} = \int \mathrm{d}x\ u_{x} \left(\partial^{-1}v\right)^{n},\qquad
\tilde{G}_{n} = \int \mathrm{d}x\ \left[u^{2}v +
  \frac{2}{2+n}\left(\partial^{-1}v\right)^{2}\right]
\left(\partial^{-1}v\right)^{n}.
\end{equation}

However, a scalar Lax representation for the system is not known which
leads to difficulties in
applying the standard techniques of supersymmetrizing the
system. Therefore, we construct directly the $N=1$ supersymmetric
extension of the non-local gas equation which is integrable as
follows.

Let us introduce the two fermionic superfields
\begin{equation}
U = \psi + \theta u,\quad V = \chi + \theta v,
\end{equation}
where $\theta$ represents the Grassmann coordinate of the
supermanifold and we assume the canonical dimensions 
\begin{equation}
[x] = -1,\quad [t] = 2,\quad \left[U\right] = \frac{1}{2},\quad
\left[V\right] = \frac{7}{2}.
\end{equation}
Writing out the most general equation of dimension
$\left[\frac{5}{2}\right]$, the one that leads to the correct bosonic
limit (and is integrable) is given by
\begin{eqnarray}
U_{t} & = & - U_{x} \left(DU\right) + \left(D^{-2}V\right),\\
V_{t} & = & - \left((DU) V\right)_{x},\label{susyequation}
\end{eqnarray}
where
\begin{equation}
D = \frac{\partial}{\partial\theta} + \theta \frac{\partial}{\partial
  x},\quad D^{2} = \partial_{x},
\end{equation}
represents the supercovariant derivative on the superspace. In
components, the equations take the forms
\begin{eqnarray}
u_{t} & = & - uu_{x} + (\partial^{-1}v),\nonumber\\
v_{t} & = & - \left(uv + \psi_{x}\chi\right)_{x},\nonumber\\
\psi_{t} & = & - \psi_{x} u + (\partial^{-1}\chi),\nonumber\\
\chi_{t} & = & - \left(u\chi\right)_{x}.
\end{eqnarray}
Clearly,
this represents a nontrivial $N=1$ supersymmetrization (and not a
$B$-supersymmetrization) of (\ref{gasequation}). However, it is not
obvious immediately if this is integrable.

We note that the system of equations (\ref{susyequation}) is a
Hamiltonian system of equations. In fact, the Hamiltonian
\begin{equation}
H_{3} = \frac{1}{2} \int \mathrm{d}Z \left(V (DU)^{2} +
\left(D^{-2}V\right)\left(D^{-1}V\right)\right),
\end{equation}
where $\mathrm{d}Z = \mathrm{d}x \mathrm{d}\theta$, together with the
Hamiltonian structure
\begin{equation}
{\cal D}_{1} = \left(\begin{array}{rcr}
0 &  & - D\\
\noalign{\vskip 2pt}%
- D &  & 0
\end{array}\right),\label{firststructure}
\end{equation}
leads to (\ref{susyequation}) as 
\begin{equation}
\left(\begin{array}{c}
U_{t}\\
V_{t}
\end{array}\right) = {\cal D}_{1} \left(\begin{array}{c}
\frac{\delta H_{3}}{\delta U}\\
\noalign{\vskip 2pt}%
\frac{\delta H_{3}}{\delta V}
\end{array}\right).
\end{equation}
Similarly, it can also be checked that (\ref{susyequation}) can also
be written as Hamiltonian equations with the Hamiltonian
\begin{equation}
H_{2} = \int \mathrm{d}Z\ U \left(DV\right),
\end{equation}
and the Hamiltonian structure
\begin{equation}
{\cal D}_{2} = \left(\begin{array}{ccc}
D^{-3} &  & -\frac{1}{2}\left(U_{x} + D^{-1} (DU_{x})\right)\\
\noalign{\vskip 2pt}%
\frac{1}{2}\left(U_{x} + (DU_{x}) D^{-1}\right) &  &  -\left(D (DV) +
(DV) D - \frac{3}{2} D V D\right)
\end{array}\right),\label{secondstructure}
\end{equation}
as
\begin{equation}
\left(\begin{array}{c}
U_{t}\\
V_{t}
\end{array}\right) = {\cal D}_{2} \left(\begin{array}{c}
\frac{\delta H_{2}}{\delta U}\\
\noalign{\vskip 2pt}%
\frac{\delta H_{2}}{\delta V}
\end{array}\right).
\end{equation}

The Hamiltonian structures (\ref{firststructure}) and
(\ref{secondstructure}) have the necessary symmetry properties and it
can be checked through the method of prolongation \cite{olver} that
they satisfy
Jacobi identity as well. Therefore, both ${\cal D}_{1}$ as well as
${\cal D}_{2}$ define genuine Hamiltonian structures. In fact, it is
obvious that ${\cal D}_{1}$ satisfies Jacobi identity trivially. It
can also be checked that the change of variables
\begin{equation}
\tilde{U} = U_{x},\quad \tilde{V} = V - \frac{1}{2} (DU_{x}) U_{x} = V
- \frac{1}{2} (D\tilde{U}) \tilde{U},\label{vtilde}
\end{equation}
diagonalizes the second Hamiltonian structure which coincides with the
supersymmetric $SL (2)\otimes U(1)$ algebra and satisfies Jacobi
identity. 
More, importantly, it can also be checked through the method of
prolongation that an arbitrary linear combination
\begin{equation}
{\cal D} = {\cal D}_{2} + \alpha {\cal D}_{1},
\end{equation}
also defines a Hamiltonian structure (satisfies Jacobi identity) so
that we conclude that the system of equations (\ref{susyequation}) is
truly a bi-Hamiltonian system. It follows now from Magri's theorem
that the supersymmetric system of equations is integrable.

From the two Hamiltonian structures in (\ref{firststructure}) and
(\ref{secondstructure}), we can obtain the recursion operator
associated with the system defined as
\begin{eqnarray}
{\cal R} & = & {\cal D}_{2} {\cal D}_{1}^{-1}\\
 & = & \left(\begin{array}{ccc}
\frac{1}{2}\left(U_{x} + D^{-1} (DU_{x})\right) D^{-1} &  &  -
 D^{-4}\\
\noalign{\vskip 2pt}%
(D^{2}V) D^{-1} + \frac{1}{2} (DV) + \frac{3}{2} V D &   &  -
 \frac{1}{2}\left(U_{x} + D^{-1}(DU_{x}\right) D^{-1}
\end{array}\right).\label{recursion}
\end{eqnarray}
This helps us determine the conserved quantities associated with the
system recursively and the first few take the forms
\begin{eqnarray}
H_{1} & = & \int \mathrm{d}Z\ V,\nonumber\\
H_{2} & = & \int \mathrm{d}Z\ U (DV),\nonumber\\
H_{3} & = & \frac{1}{2} \int \mathrm{d}Z\left[(DU)^{2} V +
  \left(D^{-1}V\right)\left(D^{-2}V\right)\right],\nonumber\\
H_{4} & = & \int \mathrm{d}Z\left[(DU)^{3} V - 3 U
  \left(D^{-1}V\right)^{2} - 6 (DU) V
  \left(D^{-3}V\right)\right],\nonumber\\
 &\vdots &  \label{nonlocalcharges}
\end{eqnarray} 
These charges are all conserved and are in involution with one another
by construction reflecting the integrability of the system.

It is worth noting that all the charges in the series
(\ref{nonlocalcharges}) are bosonic. In addition, we have found a
charge that is fermionic and is conserved under the flow. It has the form
\begin{equation}
\tilde{H} = \int \mathrm{d}Z\ UV \left(D^{-2}V\right).
\end{equation}
Furthermore, much like the bosonic system, we have also found two
series of bosonic conserved charges that are not related recursively
\begin{eqnarray}
G_{n} & = & \int \mathrm{d}Z\ U (DV) (D^{-1}V)^{n},\nonumber\\
\tilde{G}_{n} & = & \int \mathrm{d}Z\ (\partial^{-1}V)
(D^{-1}V)^{n}\left[(n+2) \left(D^{-1}V\right)^{2} + (n+1)(n+3) U_{x}V
  (DU) - (n+3) (D^{-1}U)(DU)(DU_{x})\right].
\end{eqnarray}

Much like the bosonic system, the $N=1$ supersymmetric system also has
Casimir functionals whose gradients are annihilated by the two
Hamiltonian structures of the system. It is easy to check that the
three conserved quantities
\begin{eqnarray}
H_{1} & = & \int \mathrm{d}Z\ V,\nonumber\\
H_{1}^{(1)} & = & - \int \mathrm{d}Z\ U_{x} \rightarrow 0,\nonumber\\
H_{1}^{(2)} & = & \int \mathrm{d}Z\ U,\label{casimir1}
\end{eqnarray}
are conserved and are Casimir functionals of the first Hamiltonian
structure (\ref{firststructure}), namely,
\begin{equation}
{\cal D}_{1} \left(\begin{array}{c}
\frac{\delta H_{1}}{\delta U}\\
\noalign{\vskip 2pt}%
\frac{\delta H_{1}}{\delta V}
\end{array}\right) = {\cal D}_{1} \left(\begin{array}{c}
\frac{\delta H_{1}^{(1)}}{\delta U}\\
\noalign{\vskip 2pt}%
\frac{\delta H_{1}^{(1)}}{\delta V}
\end{array}\right) = {\cal D}_{1}\left(\begin{array}{c}
\frac{\delta H_{1}^{(2)}}{\delta U}\\
\noalign{\vskip 2pt}%
\frac{\delta H_{1}^{(2)}}{\delta V}
\end{array}\right) = 0.
\end{equation}
This is very much like in the bosonic model. Furthermore, the second
Hamiltonian structure (\ref{secondstructure}) also has a Casimir
functional
\begin{equation}
H_{-1} =  \int \mathrm{d}Z\ \frac{V - \frac{1}{2} (DU_{x})
  U_{x}}{\sqrt{(DV) - \frac{1}{2} (DU_{x})^{2} - \frac{1}{2} U_{xx}
  U_{x}}},\label{casimir2}
\end{equation}
which is conserved and satisfies
\begin{equation}
{\cal D}_{2} \left(\begin{array}{c}
\frac{\delta H_{-1}}{\delta U}\\
\noalign{\vskip 2pt}%
\frac{\delta H_{-1}}{\delta V}
\end{array}\right) = 0.
\end{equation}

The existence of Casimir functionals suggests that the hierarchy of
equations generated by $H_{n}, n>0$ can be extended to the negative
orders through the inverse of the recursion operator
(\ref{recursion}). Formally, the inverse can be defined as
\begin{equation}
{\cal R}^{-1} = {\cal D}_{1} {\cal D}_{2}^{-1}.
\end{equation}
However, it is worth noting here that unlike in the bosonic case where
a closed form expression for the inverse exists, here we have not
found such a form. However, one can easily construct the conserved
charges associated with the negative order of the hierarchy using the
recursion relation
\begin{equation}
{\cal D}_{1}\left(\begin{array}{c}
\frac{\delta H_{-n}}{\delta U}\\
\noalign{\vskip 2pt}%
\frac{\delta H_{-n}}{\delta V}
\end{array}\right) = {\cal D}_{2} \left(\begin{array}{c}
\frac{\delta H_{-n-1}}{\delta U}\\
\noalign{\vskip 2pt}%
\frac{\delta H_{-n-1}}{\delta V}
\end{array}\right).
\end{equation}
For example, this leads to the first few conserved local charges of
the forms 
\begin{eqnarray}
H_{-1} & = & - \int \mathrm{d}Z\
\frac{\tilde{V}}{\sqrt{(D\tilde{V})}},\nonumber\\
H_{-2} & = & \frac{1}{2} \int
\mathrm{d}Z\left[\frac{U_{xx}}{\sqrt{(D\tilde{V})}} - 
  \frac{(DU_{xx})\tilde{V}}{(D\tilde{V})^{\frac{3}{2}}}\right],\nonumber\\
H_{-3} & = & \frac{1}{24} \int \mathrm{d}Z\left[
  \frac{\left(12 U_{xx}
    (DU_{xx}) + 10 \tilde{V}_{xx}\right)}{(D\tilde{V})^{\frac{3}{2}}}
  -  \frac{9\tilde{V}
    \left((D\tilde{V}_{xx}) +
    (DU_{xx})^{2}\right)}{(D\tilde{V})^{\frac{5}{2}}}\right],\nonumber\\
H_{-4} & = & \frac{3}{2} \int \mathrm{d}Z\left[\frac{U_{xx}
  (D\tilde{V}_{xx}) + U_{xx} (DU_{xx})^{2} - \tilde{V}_{x}
    (DU_{xx})}{(D\tilde{V})^{\frac{3}{2}}}\right.\nonumber\\
 &  & \qquad - \frac{9U_{xxx}(D\tilde{V}_{x}) - 5U_{xx}
    (D\tilde{V}_{xx}) - 5U_{xx} (DU_{xx})^{2} - 10 \tilde{V}_{xx}
    (DU_{xx}) - 6 V
    (DU_{xxxx})}{(D\tilde{V})^{\frac{5}{2}}}\nonumber\\
&  & \qquad \left. + \frac{15 \tilde{V} (D\tilde{V}_{xx}) (DU_{xx}) +
    5 \tilde{V}
    (DU_{xx})^{3}}{(D\tilde{V})^{\frac{7}{2}}}\right],\nonumber\\ 
 & \vdots & 
\end{eqnarray}
where $\tilde{V}$ is defined in (\ref{vtilde}). These charges, of
course, satisfy the recursion relation (and, therefore, are in
involution)  and reduce to
the known charges in the bosonic limit, but it is
interesting to note that in the bosonic limit, the last two terms in
$H_{-4}$ vanish.

Given the Hamiltonians in the negative hierarchy, we can obtain the
equations through the known Hamiltonian structures. We simply note
that the lowest order equation in the negative hierarchy has the form
\begin{eqnarray}
U_{t_{-1}} & = & \frac{1}{4}
D\left[\frac{2}{(D\tilde{V})^{\frac{1}{2}}} -
  \frac{3\tilde{V}\tilde{V}_{x}}{(D\tilde{V})^{\frac{5}{2}}}\right],\nonumber\\
V_{t_{-1}} & = &
\frac{1}{8}\left[\frac{-4U_{xxx}}{(D\tilde{V})^{\frac{1}{2}}} + 6
  \left(\frac{U_{xx}}{(D\tilde{V})^{\frac{1}{2}}}\right)_{x} + 2
  D\left(\frac{(DU_{xx})}{(D\tilde{V})^{\frac{1}{2}}}\right)\right.\nonumber\\
 &  & \quad \left.+ \frac{6 \tilde{V}\tilde{V}_{x}
    U_{xxx}}{(D\tilde{V})^{\frac{5}{2}}} -
    9\left(\frac{\tilde{V}\tilde{V}_{x}U_{xx}}{(D\tilde{V})^{\frac{5}{2}}}
\right)_{x} - 3 D\left(\frac{\tilde{V}\tilde{V}_{x}
  (DU_{xx})}{(D\tilde{V})^{\frac{5}{2}}}\right)\right].
\end{eqnarray}

\section{$N=2$ Supersymmetrization}

The $N=2$ supersymmetrization of the non-local gas equation can now be
obtained in a simple manner. Let us define bosonic superfields
$\overline{U},\overline{V}$ in the $N=2$ extended superspace as
\begin{equation}
\overline{U} = U_{1} + \theta_{2} U,\quad \overline{V} = V_{1} +
\theta_{2} V,
\end{equation}
where $U,V$ are the $N=1$ superfields defined in the last section
while $U_{1},V_{1}$ represent two new $N=1$ bosonic superfields. In
this extended superspace, we can define two covariant derivatives as
\begin{eqnarray}
D_{1} & = & \frac{\partial}{\partial\theta_{1}} +
\theta_{1}\frac{\partial}{\partial x},\nonumber\\ 
D_{2} & = & \frac{\partial}{\partial\theta_{2}} + \theta_{2}
\frac{\partial}{\partial x},\nonumber\\
D_{1}^{2} & = & D_{2}^{2} = \frac{\partial}{\partial x},\quad
D_{1}D_{2} + D_{2}D_{1} = 0.
\end{eqnarray}
With these, the $N=2$ supersymmetric non-local gas equation which
reduces to the $N=1$ system (\ref{susyequation}) and is integrable
takes the form
\begin{eqnarray}
\overline{U}_{t} & = & - \left(D_{1}D_{2}\overline{U}\right)
\overline{U}_{x} + \left(\partial^{-1}\overline{V}\right),\nonumber\\
\overline{V}_{t} & = & -
\left(\overline{V}\left(D_{1}D_{2}\overline{U}\right)\right)_{x} +
D_{1}D_{2}\left[\overline{V}\ \overline{U}_{x} +
  \left(D_{2}\overline{U}_{x}\right)\left(D_{1}\overline{U}_{x}\right)
  \overline{U}_{x} - \left(D_{1}D_{2}\overline{U}_{x}\right)
  \overline{U}_{x}^{2}\right].\label{susyequation2}
\end{eqnarray} 

The system of equations (\ref{susyequation2}) can be written as a
Hamiltonian system with
\begin{eqnarray}
{\cal D}_{1} & = & \left(\begin{array}{cc}
0 & D_{1}D_{2}\partial^{-1}\\
D_{1}D_{2}\partial^{-1} & 0\label{d1}
\end{array}\right),\nonumber\\
H & = & \frac{1}{6} \int
\mathrm{d}\overline{Z}\left[\left(D_{1}D_{2}\overline{U}_{x}\right)
  \overline{U}_{x}^{3} -
3  \left(D_{1}D_{2}\partial^{-1}\overline{V}\right)
  \left(2\left(D_{1}D_{2}\overline{U}\right)\overline{U}_{x} -
  \left(\partial^{-1}\overline{V}\right)\right)\right],
\end{eqnarray}
(where $\mathrm{d}\overline{Z} =
\mathrm{d}x\mathrm{d}\theta_{1}\mathrm{d}\theta_{2}$) so that
\begin{equation}
\left(\begin{array}{c}
\overline{U}_{t}\\
\noalign{\vskip 2pt}%
\overline{V}_{t}
\end{array}\right) = {\cal D}_{1} \left(\begin{array}{c}
\frac{\delta H}{\delta \overline{U}}\\
\noalign{\vskip 2pt}%
\frac{\delta H}{\delta \overline{V}}
\end{array}\right).
\end{equation}
The Hamiltonian structure clearly has the necessary anti-symmetry
properties and trivially satisfies the Jacobi identity.

The system of equations (\ref{susyequation2}) has a second Hamiltonian
description as well. It can be checked that the equations can be
written in the Hamiltonian form
\begin{equation}
\left(\begin{array}{c}
\overline{U}_{t}\\
\noalign{\vskip 2pt}%
\overline{V}_{t}
\end{array}\right) = {\cal D}_{2} \left(\begin{array}{c}
\frac{\delta \overline{H}}{\delta \overline{U}}\\
\noalign{\vskip 2pt}%
\frac{\delta \overline{H}}{\delta \overline{V}}
\end{array}\right).
\end{equation}
with the Hamiltonian
\begin{equation}
\overline{H} = \int \mathrm{d}\overline{Z}\
\left(D_{1}D_{2}\overline{U}\right)\left[\overline{V} -
  \frac{1}{2} D_{1}\left(\overline{U}_{x}
  (D_{2}\overline{U}_{x})\right)\right],
\end{equation}
and the second Hamiltonian structure ${\cal D}_{2}$ with the elements
\begin{eqnarray}
\left({\cal D}_{2}\right)_{11} & = & -
\partial^{-1}D_{1}^{-1}D_{2}^{-1},\nonumber\\
\left({\cal D}_{2}\right)_{12} & = & \frac{1}{2}\left(-2
\overline{U}_{x} + D_{1}^{-1} (D_{1}\overline{U}_{x}) + D_{2}^{-1}
(D_{2}\overline{U}_{x}) + 2
D_{1}^{-1}D_{2}^{-1}(D_{1}D_{2}\overline{U}_{x})\right),\nonumber\\
\left({\cal D}_{2}\right)_{21} & = &
\frac{1}{2}\left(2\overline{U}_{x} + (D_{1}\overline{U}_{x})D_{1}^{-1}
+ (D_{2}\overline{U}_{x}) D_{2}^{-1} - 2 (D_{1}D_{2}\overline{U}_{x})
D_{1}^{-1}D_{2}^{-1}\right),\nonumber\\
\left({\cal D}_{2}\right)_{22} & = & \frac{1}{2}\left(- 2 \partial
\overline{V} - 2 \overline{V}\partial + D_{1}\overline{V} D_{1} +
D_{2}\overline{V} D_{2} - \partial \overline{U}_{x}^{2} D_{1}D_{2} +
\overline{U}_{x}^{2} \partial D_{1} D_{2} + D_{2} \overline{U}_{x}^{2}
\partial D_{1} - D_{1}\overline{U}_{x}^{2} \partial
D_{2}\right)\nonumber\\
 &  & \qquad + (D_{1}D_{2}\overline{U}_{x}) \partial^{-1}D_{1}D_{2}
\overline{U}_{x} D_{1}D_{2}.\label{d2}
\end{eqnarray}
The second Hamiltonian structure is quite complicated and one can
check Jacobi identity through the method of prolongation. However, it
is much easier to check this through a change of variables
\begin{equation}
\tilde{\overline{U}} = \overline{U}_{x},\qquad \tilde{\overline{V}} =
\overline{V} - (D_{1}D_{2}\tilde{\overline{U}}) \tilde{\overline{U}} -
\frac{1}{2} (D_{1}\tilde{\overline{U}})(D_{2}\tilde{\overline{U}}),
\end{equation}
the second Hamiltonian structure coincides with the $N=2$ supersymmetric
generalization of the $SL(2)\otimes U(1)$ algebra and thereby
satisfies the Jacobi identity. Furthermore, the compatibility of the
two structures ${\cal D}_{1}$ and ${\cal D}_{2}$ can also be checked
in a straightforward manner through prolongation. Therefore, it
follows that the system of $N=2$ supersymmetric equations
(\ref{susyequation2}) are integrable. The infinite set of conserved
charges in involution can be constructed using the recursion
relation. However, their forms are extremely complicated and are not
very enlightening. So, we do not list them here.

The  bosonic limit of (\ref{susyequation2}) leads to a new and
interesting coupled equation that is integrable. Introducing the
notation that $u_{0},u_{1}$ represent the bosonic variables of the
superfield $\overline{U}$ and $v_{0},v_{1}$ represent the bosonic
variables of the superfield $\overline{V}$, the equations can be
written as
\begin{eqnarray}
u_{0,t} & = & - u_{1}u_{1,x} +
\left(\partial^{-1}v_{0}\right),\nonumber\\
u_{1,t} & = & u_{0,xx}u_{0,x} - u_{1,x}u_{1} +
\left(\partial^{-1}v_{1}\right),\nonumber\\
v_{0,t} & = & u_{0,xxx}u_{0,x}^{2} + u_{0,xx}^{2}u_{0,x} -
u_{1,x}^{2}u_{0,x} - v_{0,x}u_{1} + v_{1}u_{0,x},\nonumber\\
v_{1,t} & = & \left(u_{1,xx} u_{0,x}^{2} + 2 u_{1,x}u_{0,xx}u_{0,x} -
v_{0,x}u_{0,x} - v_{1}u_{1}\right)_{x}.
\end{eqnarray}
This is a new integrable system of equations and has the interesting
feature that the second Hamiltonian structure for this system does not
have any Casimir functional. As a result, the second Hamiltonian
structure for the $N=2$ supersymmetric system in 
(\ref{d2}) does not also possess any Casimir functional (although the
first structure does) and the system of equations
(\ref{susyequation2}) cannot be extended to the negative hierarchy.

\section{Summary}

In this paper we have systematically studied the supersymmetrization
of the non-local gas equation preserving integrability. We obtain the
$N=1$ supersymmetric system of equations and show that it has two
compatible Hamiltonian structures making it a bi-Hamiltonian
system. We construct the conserved charges of the system and show that
the two Hamiltonian structures possess Casimir functionals. As a
result, the system of supersymmetric equations can be extended to
negative orders and these give rise to local equations of motion. We
also construct the $N=2$ supersymmetric generalization of the
non-local gas equation and show that it is a bi-Hamiltonian system. In
the bosonic limit, this equation leads to a new coupled integrable
system of equations. Furthermore, we argue that in the $N=2$
supersymmetric case, while the first Hamiltonian structure possesses
Casimir functionals, the second does not. As a result, these equations
cannot be extended to negative orders.

\vskip .7cm

\noindent{\bf Acknowledgment}
\medskip

This work was supported in part by the US DOE Grant number DE-FG
02-91ER40685.

\end{document}